\documentclass[conference,twocolumn,10pt]{IEEEtran} 
\usepackage{epsfig}
\usepackage{color}
\usepackage{amsmath}
\usepackage{amssymb}

\newtheorem{theorem}{Theorem}

\newtheorem{lemma}{Lemma}

\newtheorem{ex}{Example}

\def\Pr{{\mathrm{Pr}}}

\def\mop{\underset{\mu_{12},\:\mu_{21}}{\longleftrightarrow}}
\def\cmop{\longleftrightarrow} 

\begin{document}
\bibliographystyle{ieeetr}


\title{Ordering Finite-State Markov Channels by\\ Mutual Information}

\author{
Andrew W. Eckford\\
Department of Computer Science and Engineering, York University\\ 
4700 Keele Street, Toronto, Ontario, Canada M3J 1P3\\
Email: aeckford@yorku.ca
}

\maketitle

\begin{abstract}
	In previous work, an ordering result was given for the symbolwise probability of error
	using general Markov channels, under iterative decoding of LDPC codes.
	In this paper, the ordering result is extended to mutual information,
	under the assumption of an iid input distribution.
	For certain channels, in which the capacity-achieving input distribution 
	is iid,
	this allows ordering of the channels by capacity.
	The complexity of analyzing general Markov channels is mitigated by this ordering,
	since it is possible to immediately determine that a
	wide class of channels, with different numbers of states,
	has a smaller mutual information than a given channel.
\end{abstract}

\section{Introduction}

A finite-state Markov channel is a channel with binary inputs,
where the instantaneous values of the channel parameters are selected by
the state of a hidden Markov chain.  
Capacity and coding were 
originally studied for these channels in \cite{gol96}.

An {\em ordering} of communication channels
may be accomplished with respect
to probability of error (for a specified code), or with respect to channel capacity.
For instance, all else being equal, 
the Gaussian channel is ordered with respect to noise variance: higher noise variance
means higher probability of error for any code, as well as lower capacity.
Such orderings are attractive to researchers, since a capacity or probability
of error result in one channel can be immediately extended and applied to
other channels that are covered by the ordering.  
The problem of ordering communication channels
can be traced back to Shannon \cite{sha58},
where a partial ordering was given for memoryless channels using
general codes.

Ordering results are particularly attractive for the analysis 
of Markov channels because of the
large size of their parameter space: $O(k^2)$ parameters
for a channel with $k$ states.  
For example, if the mutual information using some
channel $\mathsf{c}$ is known, it would be helpful 
for $\mathsf{c}$ to cast a ``shadow'' of neighboring
channels where the mutual information was known to be 
smaller (or larger).
In previous work \cite{eck03,eck07}, we obtained ordering results for 
general Markov channels with respect to symbol error under iterative 
decoding of LDPC codes, and a key feature of that work was the ability
to compare channels with different numbers of states.

The contribution of the present paper is to generalize the ordering result 
from \cite{eck03,eck07} to mutual information
{\em under an iid input distribution}.
This restriction is used for
three reasons:
\begin{itemize}
	\item Mutual information under an iid input distribution 
	is by far the most practically interesting case for communications engineers;
	\item Under some circumstances, an iid input distribution is capacity-achieving 
	\cite{gol96,mus89}; and
	\item It is very difficult to analyze Markov channels under non-iid input distributions.
\end{itemize}
The iid input distribution makes our results particularly applicable to
the achievable rates of contemporary error-control codes,
whose codewords are generally considered to simulate iid input distributions.
Furthermore, the result is more general than previous work, 
applying to the ultimate limits of any possible
code whose codewords satisfy the iid input distribution, 
rather than being relevant only to LDPC codes.

To adapt these ordering results to mutual information, different
theoretical machinery is required.  This is mostly because the ordering in
\cite{eck03,eck07} was based on symbol error, but the mutual information is
related to block error.  Using symbol error, our approach was to add or delete
certain ``side information'' until the desired structure of the decoder was achieved.
In this paper, we 
start out by using a similar approach (including a proof technique 
initially used in \cite{mus89}),
although a completely 
different method is required to prove its relevance to the present ordering.
Furthermore, to take the global nature of block errors into account, we introduce
a lemma (Lemma \ref{lem:monotonic}), related to the mutual information
of channels with piecewise-Markov segments.

The remainder of the paper is organized as follows.  In Section \ref{sec:model},
we describe our model for general finite-state Markov channels.  In Section \ref{sec:mixing},
we describe a mixing operator (previously introduced in \cite{eck03,eck07}), which allows us to 
flexibly construct 
degraded channels with larger numbers of states than the original channels.
Finally, in Section \ref{sec:result}, we state and prove our main result, and give some
discussion concerning its use.

\section{Model}
\label{sec:model}

In this paper, we will write constant scalars and vectors as $x$ and $\mathbf{x}$, respectively;
and scalar and vector random variables as $X$ and $\mathbf{X}$, respectively.  
For a random variable $X$, a realization of the random variable will 
usually be written as the
corresponding lower-case letter $x$.
We also use
bold upper-case letters to represent constant matrices, but it should be clear from the context
when we mean a matrix and when we mean a vector random variable.
Finally, for probability density functions (PDFs), such as $f_X(x)$, and
probability mass functions (PMFs), such as $p_X(x)$, we will omit the subscript when it is unambiguous
to do so, and simply write $f(x)$ and $p(x)$ for PDFs and PMFs, respectively.

Consider a channel $\mathsf{c}$ with inputs selected from an alphabet $\mathcal{X}$,
outputs selected from an alphabet $\mathcal{Y}$, 
and hidden channel states selected from an alphabet $\mathcal{S}$.
The sets $\mathcal{X}$ and $\mathcal{Y}$ could be discrete or
continuous, but $\mathcal{S}$ is always discrete and finite
for a finite-state Markov channel
(for example, $\mathcal{S} = \{1,2,\ldots,|\mathcal{S}|\}$).
Let $\mathbf{X}\in\mathcal{X}^n$,
$\mathbf{Y}\in\mathcal{Y}^n$,
and $\mathbf{S}\in\mathcal{S}^{n+1}$
represent random
vectors, consisting
of channel inputs, channel outputs, and channel states, respectively.

We assume throughout the paper that $\mathbf{S}$ forms a {\em regular}
Markov chain operating in steady state, which is {\em independent} of the
channel inputs $\mathbf{X}$.  Furthermore,
given the channel state $\mathbf{S}$, 
we assume that the channel is memoryless, i.e.,
\begin{equation}
	\label{eqn:CondIndependent}
	f(\mathbf{y}\:|\:\mathbf{s},\mathbf{x}) =
	\prod_{t=1}^n f(y_t\:|\:s_t,x_t) .
\end{equation}
When both of these properties hold, then $\mathsf{c}$ 
is called a {\em Markov channel}.
These properties exclude partial response channels from the discussion.
Furthermore, the specification that the Markov chain is regular means that
there exists a steady-state distribution for the state sequence $\mathbf{S}$,
and that the state probabilities converge to the steady-state distribution.

Combining (\ref{eqn:CondIndependent}) with the PMF of $\mathbf{S}$, we can write
\begin{equation}
	\label{eqn:CondFull}
	f(\mathbf{y},\mathbf{s}\:|\:\mathbf{x}) =
	p(s_1) \prod_{t=1}^n f(y_t\:|\:s_t,x_t) p(s_{t+1}\:|\:s_t) ,
\end{equation}
and the channel input-output relationship is given by marginalizing (\ref{eqn:CondFull})
over $\mathbf{s}$.  

From (\ref{eqn:CondFull}), the channel is fully parameterized by
specifying $p(s_{t+1}\:|\:s_t)$ and $f(y_t\:|\:s_t,x_t)$.
The values of $p(s_{t+1}\:|\:s_t)$ are commonly specified in a 
$|\mathcal{S}|\times |\mathcal{S}|$ matrix $\mathbf{P}$, known as the
{\em transition probability matrix}.  If $\mathcal{S} = [1,2,\ldots,|\mathcal{S}|]$,
then the element of $\mathbf{P}$ on the $i$th row and $j$th column is given by
\begin{displaymath}
	P_{i,j} = p_{S_{t+1}|S_t}(j \:|\: i) .
\end{displaymath}
We assume that $f(y_t\:|\:s_t,x_t)$ is drawn from a given family of channels,
where $s_t$ corresponds to a particular channel parameter for that family.
For example, if $f(y_t\:|\:s_t,x_t)$ represents a binary symmetric channel,
then each possible value of $s_t$ in $\mathcal{S}$ corresponds to an
inversion probability.  Thus, these parameters can be expressed in a vector
$\mathbf{n}$, where
\begin{displaymath}
	\mathbf{n} = [\eta_1,\eta_2,\ldots,\eta_{|\mathcal{S}|}] .
\end{displaymath}
Given the family, a Markov channel $\mathsf{c}$ is completely specified by the parameters
\begin{displaymath}
	\mathsf{c} = (\mathbf{P},\mathbf{n}) .
\end{displaymath}

\section{Degrading Markov channels}
\label{sec:mixing}

\subsection{Mixing operator}

We re-use the Markov channel mixing operator from \cite{eck03,eck07},
which is based on a proof technique from \cite{mus89}.
Let $\mathsf{c}_1 = (\mathbf{P}_1,\mathbf{n}_1)$ 
and $\mathsf{c}_2 = (\mathbf{P}_2,\mathbf{n}_2)$ represent Markov channels.  
The hidden Markov chain in each channel is implemented by a Markov state machine,
$\mathcal{M}_1$ and $\mathcal{M}_2$ for channels $\mathsf{c}_1$ and $\mathsf{c}_2$,
respectively,
consisting of the possible states in each channel, connected by their
transition probabilities.  We assume that the set of states in 
$\mathsf{c}_1$ and the set of states in $\mathsf{c}_2$ are disjoint.

We will ``mix'' these channels by allowing jumps between their
respective Markov state machines, as follows.  
Let $\mathbf{U}^{(1 \rightarrow 2)}$
and $\mathbf{U}^{(2 \rightarrow 1)}$ represent Bernoulli random vectors
of the same length as the state sequences, whose elements take values in $\{0,1\}$.
The vectors $\mathbf{U}^{(1 \rightarrow 2)}$
and $\mathbf{U}^{(2 \rightarrow 1)}$ 
are independent of the channel inputs $\mathbf{X}$.
Then the ``mixed'' state machine behaves as follows:
\begin{itemize}
	\item If the state at time $t$ is in machine $\mathcal{M}_1$, and $U_t^{(1 \rightarrow 2)} = 1$,
	then the state at time $t+1$ is in $\mathcal{M}_2$, chosen at random according to the
	steady-state probabilities of the states in $\mathcal{M}_2$, and independently of any previous state.
	\item If the state at time $t$ is in machine $\mathcal{M}_2$, and $U_t^{(2 \rightarrow 1)} = 1$,
	then the state at time $t+1$ is in $\mathcal{M}_1$, chosen at random according to the
	steady-state probabilities of the states in $\mathcal{M}_1$, and independently of any previous state.
	\item If neither of these conditions hold, then the next state is chosen randomly according to
	the Markov chain probabilities in either $\mathcal{M}_1$ (if the current state is in $\mathcal{M}_1$) 
	or $\mathcal{M}_2$ (if the current state is in $\mathcal{M}_2$).
\end{itemize}

Let $\mu_{12}$ and $\mu_{21}$ represent the probabilities
$\Pr(U^{(1 \rightarrow 2)} = 1)$ and $\Pr(U^{(2 \rightarrow 1)} = 1)$, respectively.
If $\mathbf{U}^{(1 \rightarrow 2)}$ and $\mathbf{U}^{(2 \rightarrow 1)}$ are not observed,
it is straightforward to 
show that the resulting ``mixed'' channel has a transition probability
matrix given by
\begin{displaymath}
	\mathbf{P}^\prime = 
	\left[
		\begin{array}{cc}
			(1 - \mu_{12}) \mathbf{P}_1 & 
				\mu_{12} \bar{\mathbf{P}}_2 \\
			\mu_{21} \bar{\mathbf{P}}_1 & 
				(1 - \mu_{21}) \mathbf{P}_2
		\end{array}
	\right] ,
\end{displaymath}
where
$\bar{\mathbf{P}}_1$ is a matrix with the same number of rows as $\mathbf{P}_2$ and
the same number of columns as $\mathbf{P}_1$, where each row corresponds to the
steady-state probabilities of the states in $\mathsf{c}_1$; similarly,
$\bar{\mathbf{P}}_2$ is a matrix with the same number of rows as $\mathbf{P}_1$ and
the same number of columns as $\mathbf{P}_2$, where each row corresponds to the
steady-state probabilities of the states in $\mathsf{c}_2$.
Furthermore, since the mixed state machine
contains the union of the states from the original state machines, 
which were disjoint, the new
vector of channel behaviors is given by
\begin{displaymath}
	\mathbf{n}^\prime = [\mathbf{n}_1 \: \mathbf{n}_2] .
\end{displaymath}
We use the operator $\cmop$ to represent this mixing operation.
If the channel $\mathsf{c}^\prime = (\mathbf{P}^\prime,\mathbf{n}^\prime)$ is formed in this manner
from $\mathsf{c}_1$ and $\mathsf{c}_2$, with parameters $\mu_{12}$ and 
$\mu_{21}$, we write
\begin{displaymath}
	\mathsf{c}^\prime = \left( \mathsf{c}_1 \mop \mathsf{c}_2 \right) .
\end{displaymath}

We give an example to illustrate the use of the operator, as follows.
\begin{ex}
	\label{ex:mixing}
	The {\em Gilbert-Elliott channel} \cite{mus89} is a two-state Markov channel,
	where each state corresponds to a BSC with a different inversion probability.
	Let $\mathsf{c}$ be a Gilbert-Elliott channel with parameters
	\begin{displaymath}
		\mathsf{c} = (\mathbf{P},\mathbf{n}) = 
		\left(
			\left[
				\begin{array}{cc}
					0.9 & 0.1 \\ 0.1 & 0.9
				\end{array}
			\right] ,
			[0.1, 0.3]
		\right) .
	\end{displaymath}
	Also, let $\mathsf{c}^*$ be another Gilbert-Elliott channel with
	parameters
	\begin{displaymath}
		\mathsf{c}^* = (\mathbf{P}^*,\mathbf{n}^*) = 
		\left(
			\left[
				\begin{array}{cc}
					0.9 & 0.1 \\ 0.1 & 0.9
				\end{array}
			\right] ,
			[0.18, 0.34]
		\right) .
	\end{displaymath}
	In both cases, 
	it is easy to show that the steady-state probabilities of each state are given by
	$P_1 = P_2 = 0.5$.  
	Let $\mu_{12} = \mu_{21} = 0.1$.  In this case,
	if $\mathsf{c}^\prime = (\mathsf{c} \mop \mathsf{c}^*)$, then
	\begin{displaymath}
		\mathsf{c}^\prime = 
		\left(
			\left[
				\begin{array}{cccc}
					0.81 & 0.09 & 0.05 & 0.05 \\
					0.09 & 0.81 & 0.05 & 0.05 \\
					0.05 & 0.05 & 0.81 & 0.09 \\
					0.05 & 0.05 & 0.09 & 0.81
				\end{array}
			\right] ,
			[0.1 , 0.3 , 0.18, 0.34]
		\right) .
	\end{displaymath}
\end{ex}

Notice that, if $\mathbf{U}^{(1 \rightarrow 2)}$ and $\mathbf{U}^{(2 \rightarrow 1)}$ are observed, 
then the Markov chain is divided into independent piecewise-Markov segments, with the divisions
occurring at each transition between the two state machines.  This occurs because the new state
is chosen with respect to the steady-state probabilities within the new state machine,
independently of any previous state.  Thus, if the input distribution $f(\mathbf{x})$ is iid,
the channel outputs $\mathbf{y}$ are also split into independent piecewise-hidden-Markov segments.

\subsection{Broken-chain degraded families}
\label{sec:BrokenChain}

We can form a family of degraded channels based on operations similar to $\cmop$.
Let $\mathcal{D}_{\mathsf{c}}$ represent a family of {\em broken-chain degraded channels},
degraded with respect to $\mathsf{c}$, defined as follows.
For all $\mathsf{c}^* \in \mathcal{D}_{\mathsf{c}}$, there exists a 
(vector) random variable $\mathbf{U}$ with the following properties:
\begin{itemize}
	\item 
	if $\mathbf{U}$ is unknown,
	then the channel is a Markov channel with parameters $\mathsf{c}^*$;
	\item if $\mathbf{U}$ is known, then the channel is a piecewise-Markov channel,
	where each segment has parameters $\mathsf{c}$; and
	\item $\mathbf{U}$ is always independent of the channel inputs $\mathbf{X}$.
\end{itemize}

Furthermore, it is easy to see that the definition of $\mathcal{D}_{\mathsf{c}}$ is
intended to be used with the operator $\cmop$, since that operator generates channels
which are piecewise-Markov (although the parameters on those segments might be different).

We make a few remarks on this definition.  Firstly, it is quite easy to see that
$\mathsf{c} \in \mathcal{D}_{\mathsf{c}}$, since $\mathbf{U}$ can be empty.  
Secondly, if $\mathsf{c}^* \in \mathcal{D}_{\mathsf{c}}$, and we form $\mathcal{D}_{\mathsf{c}^*}$,
which is the degraded family of $\mathsf{c}^*$, then $\mathcal{D}_{\mathsf{c}^*} \subseteq 
\mathcal{D}_{\mathsf{c}}$, since the random vectors $\mathbf{U}$ can be concatenated.
Thirdly, the random variable $\mathbf{U}$ need not 
necessarily break the Markov chain -- if the Markov chain remains in one piece
and remains Markov, it is trivially piecewise-Markov.  For instance,
if $\mathsf{c}$ is a Gilbert-Elliott channel, and $\mathsf{c}^*$ is 
a channel formed by concatenating a channel having parameters $\mathsf{c}$ 
with an independent BSC, then $\mathbf{U}$ could be the independent BSC's noise sequence,
which restores the original channel.

\section{Main result}
\label{sec:result}

\subsection{Definitions and notation}

We briefly describe some important definitions and notation in this section.  
Recall that we are restricting ourselves to {\em regular Markov chains} (which is 
implicit in the term {\em Markov channel}), and 
{\em iid input distributions}.
Firstly,
since Markov channels have memory, the mutual information is defined as
\begin{displaymath}
	I(X;Y) = \lim_{n \rightarrow \infty} \frac{1}{n}I(\mathbf{X};\mathbf{Y}) ,
\end{displaymath}
where $I(\mathbf{X};\mathbf{Y})$ represents the mutual information between
the length-$n$ vector random variables $\mathbf{X}$ and $\mathbf{Y}$, and
noting that the limit exists thanks to the restrictions we have imposed.
Since we consider mutual information under various channel assumptions, we
will write
\begin{displaymath}
	I[\mathsf{c}](X;Y)
\end{displaymath}
to represent the mutual information in channel $\mathsf{c}$.
Similarly, for
the vector version, we will write
$I[\mathsf{c}](\mathbf{X};\mathbf{Y})$.
A segment of one of these vectors, for example from the $i$th element to the $j$th
element, $j > i$, is written
\begin{displaymath}
	\mathbf{x}_i^j = [x_i,x_{i+1},\ldots,x_{j-1},x_j] .
\end{displaymath}
We will write $I[\mathsf{c}](\mathbf{X}_i^j;\mathbf{Y}_i^j)$ to represent
the mutual information between these vector segments.

\subsection{Result}

The main result of this paper is stated in the following theorem.

\begin{theorem}
	\label{thm:main}
	Let $\mathsf{c}$ represent a Markov channel, 
	and let $\mathcal{D}_{\mathsf{c}}$ represent
	its degraded family.  Suppose the input distribution is iid.
	If $\mathsf{c}^* \in \mathcal{D}_{\mathsf{c}}$, then
	\begin{equation}
		\label{eqn:Theorem1}
		I[\mathsf{c}](X;Y) \geq I \left[ \mathsf{c}\mop\mathsf{c}^* \right] (X;Y) 
	\end{equation}
	for all $0 < \mu_{12} < 1$, $0 < \mu_{21} < 1$.
\end{theorem}


To prove the Theorem, we first require the following useful lemma:

\begin{lemma}
	\label{lem:monotonic}
	Let $\mathsf{c}$ represent a Markov channel.
	Then, if the input distribution is iid, 
	\begin{displaymath}
		I[\mathsf{c}](X;Y) \geq \frac{1}{k} I[\mathsf{c}](\mathbf{X}_1^k;\mathbf{Y}_1^k)
	\end{displaymath}
	for any $k < \infty$.
\end{lemma}
The lemma states that 
observing a truncated Markov channel {\em never} gives more mutual information
than a Markov channel observed over an asymptotically long period of time.
The proof for the Lemma is contained in Appendix \ref{apx:ProofMonotonic}.

We also require the following lemma, which relates the degraded family $\mathcal{D}_{\mathsf{c}}$
to the operator $\cmop$:

\begin{lemma}
	\label{lem:degraded}
	If $\mathsf{c}^* \in \mathcal{D}_{\mathsf{c}}$, then
	\begin{displaymath}
		\mathsf{c} \cmop \mathsf{c}^* \in \mathcal{D}_{\mathsf{c}} .
	\end{displaymath}
\end{lemma}
\begin{proof}
	For channel $\mathsf{c} \cmop \mathsf{c}^*$, knowledge of $\mathbf{U}^{(1 \rightarrow 2)}$
	and $\mathbf{U}^{(2 \rightarrow 1)}$ breaks the channel into piecewise segments
	of $\mathsf{c}$ and $\mathsf{c}^*$.  However, since $\mathsf{c}^* \in \mathcal{D}_{\mathsf{c}}$,
	there exists $\mathbf{U}$ to transform each piecewise segment in $\mathsf{c}^*$ to a 
	segment in $\mathsf{c}$.  Taken together, $\mathbf{U}^{(1 \rightarrow 2)}$,
	$\mathbf{U}^{(2 \rightarrow 1)}$, and $\mathbf{U}$ transform $ \mathsf{c} \cmop \mathsf{c}^*$
	into piecewise segments of $\mathsf{c}$, which is the definition of a channel in 
	$\mathcal{D}_{\mathsf{c}}$.
\end{proof}

The proof of Theorem \ref{thm:main} is then given as follows.

\begin{proof}
	Let $\mathsf{c}^\prime = \mathsf{c}\mop\mathsf{c}^*$.
	By Lemma \ref{lem:degraded}, $\mathsf{c}^\prime \in \mathcal{D}_{\mathsf{c}}$,
	so there exists a random variable $\mathbf{U}$ which transforms
	$\mathsf{c}^\prime$ into piecewise-Markov segments of $\mathsf{c}$.
	
	Let $\mathcal{J}$ represent an index set corresponding to the independent
	segments, let the subscript $i,j$ represent the $i$th symbol in the $j$th
	segment, and let $\ell(j)$ represent the length of the $j$th segment.
	Then we have that
	\begin{eqnarray*}
		\lefteqn{f(\mathbf{y}\:|\:\mathbf{x},\mathbf{u}) =} & & \\
		& & 
		\prod_{j \in \mathcal{J}} \left[ p(s_{1,j})\prod_{i=1}^{\ell(j)} f(y_{i,j}\:|\:s_{i,j},x_{i,j})
			\prod_{i=1}^{\ell(j)-1} p(s_{i+1,j}\:|\:s_{i,j}) \right] .
	\end{eqnarray*}
	Since $\mathbf{X}$ is iid (by assumption), then
	\begin{eqnarray*}
		\lefteqn{f(\mathbf{y}\:|\:\mathbf{u}) =} & & \\
		& & 
		\prod_{j \in \mathcal{J}} \left[ p(s_{1,j})\prod_{i=1}^{\ell(j)} f(y_{i,j}\:|\:s_{i,j})
			\prod_{i=1}^{\ell(j)-1} p(s_{i+1,j}\:|\:s_{i,j}) \right] ,
	\end{eqnarray*}
	which is accomplished by marginalizing over each $x_{i,j}$.
	Thus, since both 
	$f(\mathbf{y}\:|\:\mathbf{x},\mathbf{u})$
	and $f(\mathbf{y}\:|\:\mathbf{u})$
	are partitioned into independent segments, we can write
	\begin{eqnarray}
		\nonumber
		\lefteqn{I[\mathsf{c}^\prime](\mathbf{X};\mathbf{Y}\:|\:\mathbf{U}) =} & & \\
		\label{eqn:SumInfo}
		& & 
		E
		\left[
		\sum_{j \in \mathcal{J}} 
		I[\mathsf{c}](\mathbf{X}_{1,j}^{\ell(j),j};\mathbf{Y}_{1,j}^{\ell(j),j})
		\right] ,
	\end{eqnarray}
	where the expectation is taken over $\mathcal{J}$ and $\ell(j)$, which are functions of
	the random variables $\mathbf{U}$.

	Since $\mathbf{U}$ is independent of 
	$\mathbf{X}$, it is true that
	\begin{eqnarray}
		\nonumber
		I[\mathsf{c}^\prime](\mathbf{X};\mathbf{Y}) & \leq &
		I[\mathsf{c}^\prime]
		(\mathbf{X};\mathbf{Y}, \mathbf{U}) \\
		\nonumber
		& = & I[\mathsf{c}^\prime]
		(\mathbf{X};\mathbf{Y}\:|\: \mathbf{U}) + I[\mathsf{c}^\prime](\mathbf{X}; \mathbf{U}) \\
		\label{eqn:InfoIneq}
		& = & I[\mathsf{c}^\prime]
		(\mathbf{X};\mathbf{Y}\:|\: \mathbf{U}) .
	\end{eqnarray}
	Because the distribution of $\mathbf{X}$ is iid, 
	we can rewrite (\ref{eqn:SumInfo}) as
	%
	%
	%
	\begin{displaymath}
		I[\mathsf{c}^\prime](\mathbf{X};\mathbf{Y}\:|\:\mathbf{U}) = 
		E
		\left[
		\sum_{j \in \mathcal{J}} 
		I[\mathsf{c}](\mathbf{X}_1^{\ell(j)};\mathbf{Y}_1^{\ell(j)})
		\right] .
	\end{displaymath}
	From Lemma \ref{lem:monotonic}, we have that
	\begin{eqnarray}
		\nonumber
		E
		\left[
			\sum_{j \in \mathcal{J}} 
			I[\mathsf{c}^\prime](\mathbf{X}_1^{\ell(j)};\mathbf{Y}_1^{\ell(j)})
		\right] 
		& \leq &
		E
		\left[
			\sum_{j \in \mathcal{J}} \ell(j) I[\mathsf{c}](X;Y)
		\right] \\
		\label{eqn:InfoIneq2}
		& = &
		n I[\mathsf{c}](X;Y) ,
	\end{eqnarray}
	where the first inequality follows from the fact that each term under the 
	expectation on the left is less than each term under the expectation on the right, 
	and the 
	last equality follows from the fact that the sum of the lengths $\ell(j)$ 
	of all the segments equal the length $n$ of the sequence, regardless of how
	the sequence is divided.

	From (\ref{eqn:InfoIneq}), (\ref{eqn:InfoIneq2}), 
	and the definition of $I[\mathsf{c}^\prime](X;Y)$,
	we have that
	\begin{eqnarray*}
		I[\mathsf{c}^\prime](X;Y) & = & \frac{1}{n}I[\mathsf{c}^\prime](\mathbf{X};\mathbf{Y})\\
		& \leq & \frac{1}{n} I[\mathsf{c}^\prime]
		(\mathbf{X};\mathbf{Y}\:|\: \mathbf{U}) \\
		& \leq & I[\mathsf{c}](X;Y) ,
	\end{eqnarray*}
	which proves the theorem.
\end{proof}

Notice, from Lemma \ref{lem:degraded}, that the channel $\mathsf{c} \mop \mathsf{c}^*$ goes back 
into the degraded family $\mathcal{D}_{\mathsf{c}}$.  Thus, the ordering given in Theorem
\ref{thm:main} can be applied recursively to create an ordering of arbitrary size.

\subsection{Discussion}

To illustrate the use of Theorem \ref{thm:main}, we 
can expand Example \ref{ex:mixing}.
As we mentioned in Section \ref{sec:BrokenChain}, 
if $\mathsf{c}$ is a Gilbert-Elliott channel, then
one member of $\mathcal{D}_{\mathsf{c}}$ is $\mathsf{c}$ concatenated
with an independent BSC, where $\mathbf{U}$ is the noise sequence of the BSC.
Thus, we have the following:
\begin{ex}
	\label{ex:MainEx}
	From Example \ref{ex:mixing},
	it is straightforward to show that
	$\mathsf{c}^*$ is formed by concatenating $\mathsf{c}$ by
	a BSC with inversion probability $p = 0.1$.
	Thus, $\mathsf{c}^* \in \mathcal{D}_{\mathsf{c}}$, and applying Theorem \ref{thm:main},
	it is true that
	$I[\mathsf{c}](X;Y) \geq I[\mathsf{c}^\prime](X;Y)$.

	Furthermore, notice that the ordering can now be applied recursively: 
	by combining $\mathsf{c}^\prime$ with $\mathsf{c}$ 
	(and optionally concatenating $\mathsf{c}^\prime$ 
	with a BSC), we obtain a channel with six states, which is degraded with respect to $\mathsf{c}$;
	continuing the process, we can obtain degraded channels with eight states, ten states, and so on,
	each time adding $\mu_{12}$ and $\mu_{21}$ as degrees of freedom.
\end{ex}


The assumption that the input density $p(\mathbf{x})$ is iid is critical to
our analysis.  Unfortunately, as noted in \cite{gol96}, it is frequently difficult
to prove capacity results for Markov channels with non-iid inputs (although
general capacity results were given in \cite{hol03}, using Lyapunov exponents).  
We leave to future work
the open problem of extending of our ordering to channels with general inputs.

\section{Acknowledgments}

The author wishes to acknowledge a stimulating discussion
with Prof. Frank R. Kschischang, of the University of Toronto,
that led him to pursue this problem.

\appendix

\subsection{Proof of Lemma \ref{lem:monotonic}}
\label{apx:ProofMonotonic}

We give the proof for discrete-valued 
random variables $\mathbf{X},\mathbf{Y}$.  It is straightforward
to generalize the proof to the case of continuous-valued random variables,
and we describe how to do so at the end.

For convenience, suppose that there exists an integer $h$ such that $hk = n$.
Let the vector $\mathbf{x}$ be broken up into segments of length $k$,
so that
\begin{eqnarray*}
	\mathbf{x} & = & [\mathbf{x}_{1}^k,\mathbf{x}_{k+1}^{2k},\ldots,\mathbf{x}_{(h-1)k + 1}^{hk}] \\
	& = & [\mathbf{x}^{(1)},\mathbf{x}^{(2)},\ldots,\mathbf{x}^{(h)}],
\end{eqnarray*}
where we use $\mathbf{x}^{(i)}$ to represent $\mathbf{x}_{(i-1)k+1}^{ik}$.  Similarly,
the vector $\mathbf{y}$ is represented by
\begin{displaymath}
	\mathbf{y} = [\mathbf{y}^{(1)},\mathbf{y}^{(2)},\ldots,\mathbf{y}^{(h)}] .
\end{displaymath}

Suppose that, after transmitting $\mathbf{x}^{(i)}$, the transmitter waits (and does not
transmit) for $(d-1)k$ channel uses, for some very large integer $d$, before transmitting 
$\mathbf{x}^{(i+1)}$.  Since (by assumption) the Markov chain is regular,
the transition probability matrix between 
the received value $y_k^{(i)}$ and $y_1^{(i+1)}$ is 
(almost) given by
\begin{displaymath}
	p(s_1^{(i+1)}\:|\:s_k^{(i)}) \approx p(s_1^{(i+1)}) .
\end{displaymath}
In fact, let $\delta$ represent the maximum deviation from $p(s_1^{(i+1)})$,
so that
\begin{equation}
	\label{eqn:deviation}
	p(s_1^{(i+1)}) (1 - \delta) 
	\leq 
	p(s_1^{(i+1)}\:|\:s_k^{(i)}) 
	\leq 
	p(s_1^{(i+1)}) (1 + \delta) 
\end{equation}
for all $s_1^{(i+1)}$ and $s_k^{(i)}$.
If the Markov chain is regular, it is well known that 
$\delta \rightarrow 0$ as $d \rightarrow \infty$.

Then we have that 
\begin{equation}
	\label{eqn:UpBound}
	f(\mathbf{y}\:|\:\mathbf{x}) \leq (1+\delta)^h 
	\prod_{i=1}^h f(\mathbf{y}^{(i)}\:|\:\mathbf{x}^{(i)})
\end{equation}
and
\begin{equation}
	\label{eqn:LowBound}
	f(\mathbf{y}\:|\:\mathbf{x}) \geq (1-\delta)^h 
	\prod_{i=1}^h f(\mathbf{y}^{(i)}\:|\:\mathbf{x}^{(i)}) .
\end{equation}
Let $\epsilon^+ = (1+\delta)^h$, and let $\epsilon^- = (1-\delta)^h$.
Calculating $H(\mathbf{Y}\:|\:\mathbf{X})$, we have that
\begin{eqnarray}
	\nonumber
	\lefteqn{H(\mathbf{Y}\:|\:\mathbf{X})} & & \\
	\nonumber
	& = & - \int_{\mathbf{x},\mathbf{y}} 
		f(\mathbf{x},\mathbf{y}) \log f(\mathbf{y}\:|\:\mathbf{x}) \\
	\label{eqn:Entropy1}
	& = & - \int_{\mathbf{x}} p(\mathbf{x})
		\int_{\mathbf{y}} f(\mathbf{y}\:|\:\mathbf{x}) \log f(\mathbf{y}\:|\:\mathbf{x}) .
\end{eqnarray}
However, from the bounds above, we can write
\begin{eqnarray}
	\nonumber
	\lefteqn{\int_{\mathbf{y}} f(\mathbf{y}\:|\:\mathbf{x}) \log f(\mathbf{y}\:|\:\mathbf{x})} & & \\
	\nonumber
	& \leq & - \int_{\mathbf{y}} \epsilon^+ \prod_{i=1}^h 
		f(\mathbf{y}^{(i)}\:|\:\mathbf{x}^{(i)})  
		\log \epsilon^- \prod_{i=1}^h p(\mathbf{y}^{(i)}\:|\:\mathbf{x}^{(i)}) \\
	\nonumber
	& = & - \epsilon^+ \sum_{i=1}^h \int_{\mathbf{y}^{(i)}} 
		f(\mathbf{y}^{(i)}\:|\:\mathbf{x}^{(i)}) \log f(\mathbf{y}^{(i)}\:|\:\mathbf{x}^{(i)}) \\
	\label{eqn:Entropy2}
	& & - \: \epsilon^+ \log \epsilon^- ,
\end{eqnarray}
where the inequality follows from the fact that $\log p(\mathbf{y}^{(i)}\:|\:\mathbf{x}^{(i)})$ is
always negative.  Combining (\ref{eqn:Entropy1}) and (\ref{eqn:Entropy2}), 
and recalling that $p(\mathbf{x})$ is iid, we have that
\begin{eqnarray*}
	\lefteqn{H(\mathbf{Y}\:|\:\mathbf{X}) =} & & \\
	& = & \epsilon^+ \sum_{i=1}^h H(\mathbf{Y}^{(i)}\:|\:\mathbf{X}^{(i)}) - \epsilon^+ \log \epsilon^- \\
	& = & \epsilon^+ h H(\mathbf{Y}_1^k\:|\:\mathbf{X}_1^k) - \epsilon^+ \log \epsilon^- .
\end{eqnarray*}
Similarly, it can be shown that
\begin{displaymath}
	H(\mathbf{Y}\:|\:\mathbf{X}) 
	\geq 
	\epsilon^- h H(\mathbf{Y}_1^k\:|\:\mathbf{X}_1^k) - \epsilon^- \log \epsilon^+ .
\end{displaymath}
Furthermore, since $\mathbf{X}$ is iid, then marginalizing (\ref{eqn:UpBound}) and (\ref{eqn:LowBound})
with respect to $\mathbf{x}$, and following the same derivation, results in
\begin{displaymath}
	H(\mathbf{Y}) \leq \epsilon^+ h H(\mathbf{Y}_1^k) - \epsilon^+ \log \epsilon^- 
\end{displaymath}
and
\begin{displaymath}
	H(\mathbf{Y}) \geq \epsilon^- h H(\mathbf{Y}_1^k) - \epsilon^- \log \epsilon^+ .
\end{displaymath}
Thus, $\lim_{d \rightarrow \infty} H(\mathbf{Y}\:|\:\mathbf{X}) = h H(\mathbf{Y}_1^k\:|\:\mathbf{X}_1^k)$
and $\lim_{d \rightarrow \infty} H(\mathbf{Y}) = h H(\mathbf{Y}_1^k)$, so 
\begin{displaymath}
	\lim_{d \rightarrow \infty} I(\mathbf{X};\mathbf{Y}) = h (H(\mathbf{Y}_1^k) -  
	H(\mathbf{Y}_1^k\:|\:\mathbf{X}_1^k) ).
\end{displaymath}
Thus, the average information rate of this channel is given by
\begin{displaymath}
	\lim_{d \rightarrow \infty} \frac{1}{hdk} I(\mathbf{X};\mathbf{Y}) = \frac{1}{dk} 
(H(\mathbf{Y}_1^k) -  
	H(\mathbf{Y}_1^k\:|\:\mathbf{X}_1^k) ) ,
\end{displaymath}
where $hdk$ is the total number of channel uses.

Notice that there are $(d-1)k$ channel uses left unused for every $k$ that are used.  We can fill these
using the same method, transmitting for $k$ channel uses and waiting (and {\em ignoring the channel}) for
$(d-1)k$ channel uses.  In this case, the total information rate improves by a factor of $d$, to
\begin{displaymath}
	\lim_{d \rightarrow \infty} \frac{1}{hdk} I(\mathbf{X};\mathbf{Y}) = \frac{1}{k} 
	(H(\mathbf{Y}_1^k) - H(\mathbf{Y}_1^k\:|\:\mathbf{X}_1^k) ) ,
\end{displaymath}
However, it is obvious that the channel capacity in this case is given by $I(X;Y)$.  Thus,
by the data processing inequality,
\begin{displaymath}
	I(X;Y) \geq  \frac{1}{k} 
	(H(\mathbf{Y}_1^k) - H(\mathbf{Y}_1^k\:|\:\mathbf{X}_1^k) ) ,
\end{displaymath}
and the lemma follows.

To generalize the lemma to continuous-valued $\mathbf{X},\mathbf{Y}$, it is necessary to take
into account the fact that $\log f(\mathbf{y}\:|\:\mathbf{x})$ could be positive.  In this case,
the inequality leading up to (\ref{eqn:Entropy2}) is broken up into integrals over
which $\log f(\mathbf{y}\:|\:\mathbf{x})$ is positive and negative, and the appropriate bound
is used over both regions; the result is a mixture of the two bounds given above, so the
convergence result holds.


\end{document}